\documentclass[letterpaper,12pt,showpacs,preprintnumbers,amsmath,amssymb]{revtex4}
\usepackage{graphicx,epsfig,graphics,amssymb,amsmath}
\begin{document}
\def \be{\begin{equation}}
\def \ee{\end{equation}}
\def \bea{\begin{eqnarray}}
\def \eea{\end{eqnarray}}
\def\beps{\mbox{\boldmath $\mathbf{\epsilon}$ }    }
\def\bxi{\mbox{\boldmath $\mathbf{\xi}$ }    }
\def\bbeta{\mbox{\boldmath $\mathbf{\eta}$ }    }
\def\bx{{\bf x}}
\def\cN{{\cal N}}
\def\by{{\bf y}}
\def\bs{{\bf s}}
\def\br{{\bf r}}
\def\bE{{\bf E}}
\def\bA{{\bf A}}
\def\bD{{\bf D}}
\def\bq{{\bf q}}
\def\bd{{\bf d}}
\def\bn{{\bf n}}
\def\bl{{\bf l}}
\def\bj{{\bf j}}
\def\bz{{\bf z}}
\def\cO{{\cal O}}

\title{Domain Patterns in the Microwave-Induced Zero-Resistance State}
\author{Ilya Finkler$^1$, Bertrand I. Halperin$^1$, Assa Auerbach$^2$,
and Amir Yacoby$^3$,}
\affiliation{
$^1$Physics Laboratories, Harvard University, Cambridge, MA 02138,USA.\\
$^2$Physics Department, Technion, Haifa 32000, Israel.\\
$^3$Department of Physics, Weizmann Institute of Science, Rehovot, Israel.}
\date{\today}
\begin{abstract}
It has been proposed that the microwave-induced ``zero-resistance'' phenomenon, observed
in a GaAs two-dimensional electron system at low temperatures in
moderate magnetic fields, results from a state with multiple domains,
in which a  large local electric field $\bE(\br)$ is oriented in
different directions. We explore here the questions of what may
determine the domain arrangement in a given sample, what do the
domains look like in representative cases, and what may be
the consequences of domain-wall localization on the macroscopic dc
conductance. We consider both effects of sample boundaries and effects
of disorder, in a simple model, which has a constant Hall
conductivity, and is characterized by a Lyapunov functional. 
\end{abstract}
 \maketitle

\section{Introduction}

The spontaneous formation of domain patterns is a frequent occurrence
in non-equilbirium systems, and it has been extensively studied in
fluid systems subject to heating or rotation, in non-equilibrium
crystal growth, in driven  interfaces between immiscible fluids, in
roughness patterns on fracture surfaces,  in chemically reacting
fluids, in dielectric breakdown, in liquid crystal structures, in
nonlinear optics, and in motion of granular materials
\cite{gollub99,cross, langer80}.
In some cases, domain formation is truly an example of spontaneously
broken symmetry; i.e., there are two or more domain structures which
are equally possible, and selection between them is triggered by
thermal noise or some other accident of the history of the sample.  In
other cases, small perturbations due to frozen-in disorder,
peculiarities at the boundaries, or other deviations of the physical
system from a symmetric idealized model are responsible.  In this
paper, we discuss a recently discovered system  where spontaneous
formation of domains is believed to occur and to play an important
role in  macroscopically observable properties, the so-called
microwave-induced zero-resistance state in two-dimensional electron
systems, in the presence of a moderately-strong magnetic field, at low
temperatures.  

Experiments on very high mobility two-dimensional electron systems in
 GaAs heterostructures,  in fields ranging from about ten to a few
 hundred millitesla,  have shown large changes in the dc resistance in
 the presence of microwave radiation, whose sign depends on the ratio
 $\tilde{\omega} \equiv \omega / \omega_c$ between the microwave
 frequency $\omega$ and the cyclotron frequency $\omega_c$\cite{ZRS-exp}.
 In particular, decreases in the resistance have been observed when
 $\tilde{\omega}$ is somewhat larger than the nearest integer.
 Moreover, in these frequency intervals, if the microwave power is
 sufficiently high, the resistance can drop by several orders of
 magnitude, perhaps falling below the experimental sensitivity, whence
 the designation Zero Resistance State.  We note that in the
 presence of a magnetic field, where there is a nonzero Hall
 conductance, zero longitudinal resistivity is equivalent to zero
 longitudinal conductivity: the current is perpendicular to the
 electric field. Here, we find it more convenient to emphasize the
 conductivity rather than resistivity.  The longitudinal resistance is
 most directly measured in a Hall bar geometry, whereas the
 longitudinal conductance is measured in an annular Corbino geometry,
 illustrated in Fig.~\ref{corbino} below. 
  
Two distinct microscopic mechanisms for conductivity corrections have been proposed:
(i) the displacement photocurrent (DP)\cite{ZRS-DP},  which is caused by
photoexcitation of electrons into displaced guiding centers; and (ii)
the distribution function  (DF) mechanism,  which involves redistribution of
intra Landau level population for large inelastic lifetimes\cite{ZRS-DF}.

Andreev {\em et. al.}\cite{ZRS-macro}, have noted that irrespective of
microscopic details, once the radiation is strong enough to render the
{\em local} conductivity negative, the system as a whole will break
into domains of photogenerated fields and Hall currents. In the
proposed domain phase, motion of domain walls can accommodate the
external voltage, resulting in  a Zero Conductance State (ZCS) in the
Corbino geometry, or a Zero Resistance State for the Hall bar
geometry, in apparent agreement with experimental reports\cite{ZRS-exp}.

Nevertheless, many questions remain to be answered.  What determines
the arrangement of domains and domain walls in an actual sample?  Is
the domain pattern static, or does it evolve periodically or
chaotically in time?  What are the effects of sample boundaries and
inhomogeneities?  What do the domain patterns look like in
representative cases?  If there are favored positions for the domain walls,
what effects will this have on the measured electrical conductance?

We address some of these questions here in the framework of a simple
phenomenological model. Further details  may be found in a previous
publication by the current authors.\cite{prl}

\section{Model.}
\subsection{General form}
In our model,  in the presence of the microwave field,  there is a local, non-linear relation 
 between the dc current $\bj(\br)$ and the local electrostatic field $\bE(\br)$:
\be \bj=\bj^{d}(\bE,\br)+ \bj^H(\bE,\br) ,
\label{curr}\ee
where $\bj^H$ is the Hall current and $\bj^d$ is the dissipative  current, defined by the requirements $\sigma^d_{\alpha \beta} = \sigma^d_{ \beta \alpha}$, 
$\sigma^H_{\alpha \beta} = - \sigma^H_{ \beta \alpha}$, 
where 
\be \sigma^d_{\alpha \beta}(\bE) \equiv \partial j^d_\alpha / \partial E_\beta .
\label{jdiss}
\ee
\be \sigma^H_{\alpha \beta}(\bE) \equiv \partial j^H_\alpha / \partial E_\beta .
\label{jhall}
\ee
These equations must be combined with the continuity equation, 
$\nabla \cdot \bj= -\dot{\rho}$,
where $\rho$ is the charge density, and additional  equations that
relate the electric field to the charge density. 

In our simplest model, we assume that the Hall current is given by a linear relation
\be \bj^{H}=\sigma^H \hat{\bz}\times \bE \  ,
\label{hall}\ee
and the Hall conductance $\sigma^H$ is assumed to be a constant,
independent of position. As we shall see below, there exists a {\em
  Lyapunov functional}\cite{cross}, which we can use to determine
the steady states, the global conductance and stability conditions on
domain walls in the strong radiation regime.

The function $\bj^d$, in general, will depend explicitly on the
position $\br$, due, e.g., to inhomogeneities in the 2D electron
system, and the direction of $\bj^d$ may not be perfectly aligned with
$\bE$. Equation (\ref{curr}) is valid above an ultraviolet cutoff,
which for the system under consideration is probably of the order of
the cyclotron radius, of order 1 $\mu$m.

Writing $\bE \equiv - \nabla \phi$, we may relate changes in the
electrostatic potential $\phi$ to changes 
in $\rho$ through the inverse capacitance matrix $W$: \be \delta
\phi(\br)=\int d^2r W(\br,\br') \delta \rho(\br'). \label{ES} \ee If a
time-independent steady state is reached, then we have simply $\nabla
\cdot \bj=0$, and the precise form of $W$ is unimportant.

In a Corbino geometry, one specifies the potential on the inner and
outer boundaries of the sample, and one looks for a solution for
$\phi(\br)$ consistent with these boundary conditions. If $\sigma^H$
is a constant, then the Hall current cannot contribute to $\nabla
\cdot \bj$ in the interior of the sample, so it does not appear in
Kirchoff's equations. Consequently, the solution for $\phi(\br)$ is
independent of $\sigma^H$, and we may, for simplicity, set
$\sigma^H=0$. To recover the Hall current, one inserts the solution
$\bE$ into the second term in (\ref{curr}).
 
The symmetry  condition on $\sigma^d_{\alpha\beta}$ allows us  to define a
Lyapunov functional for $\bj^d$, as 
\be
G[\phi]= \int d^2x g(\bE),~~~g= \int_0^{\bE(\bx)} d\bE' \cdot \bj^d(\bE')
\label{G-def}
\ee
A variation of (\ref{G-def}) is given by
\be
\delta G =\int d^2x  \nabla \cdot \bj^d~\delta \phi
-\int_{bound}ds\hat{\bn} \cdot \bj^d \delta\phi \label{dG} \ee
The second integral vanishes on equipotential boundaries, or in the
absence of external currents.  The extrema of $G$ are found to be
steady states, with  $\nabla \cdot \bj=0$. Using the positivity of the
inverse capacitance matrix $W$, one may show  that $G[\phi(t)]$ is
indeed a Lyapunov functional, i.e. a non-increasing function of time,
so that its minima are stable steady states. In general, $G$ may have
multiple minima. Any initial choice of $\phi(\br)$ will relax to some
local minimum of $G$, but not necessarily the ``ground state'' with
lowest $G$.  Nevertheless, we might expect that the system is most
likely to wind up in a state with $G$ close to the absolute minimum,
and in the presence of noise, we might expect that the system might
tend to escape from high-lying minima.

Using the boundary term in (\ref{dG}), the current across a Corbino
sample is equal to the first derivative of $G$ with respect to the
potential difference $\delta V$ between two edges, and the conductance
is given by the stiffness, or the second derivative:
\be \frac{\delta I}{\delta V} = \frac{d^2 G}{d(\delta V)^2}.
\label{conduct} \ee

\subsection{Strong Radiation.}

\begin{figure}[!h]
\vspace{1mm}
\centerline{\epsfysize=2in\epsfbox{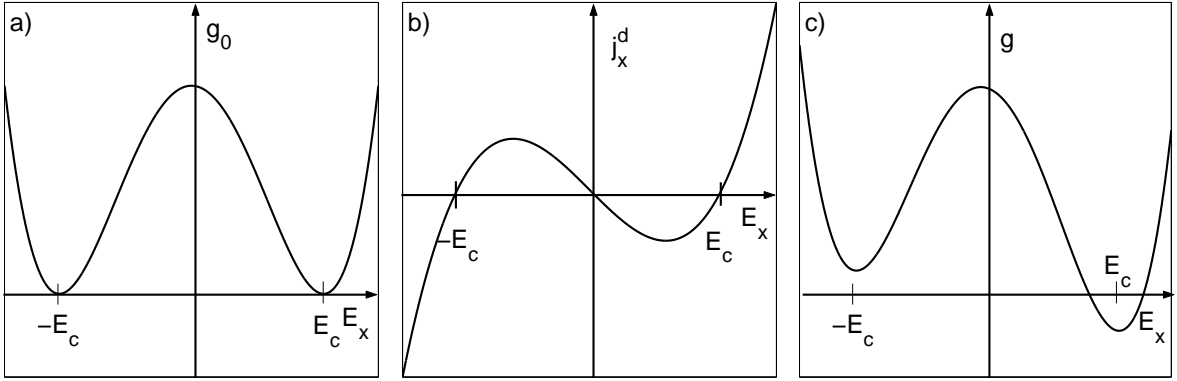}}
\vspace{1mm}
\caption{Model Lyapunov functions $g$ and dissipative current $j^d_x$
  for a uniform electrostatic field $E_x$ along the $x$-axis
  ($E_y=0$), in the presence of strong microwave irradiation.
(a) Lyapunov  function $g_0(E_x)$ in the absence of disorder. Operating points are at the 
minima of g, where $E_x= \pm E_c$.
(b) Dissipative current corresponding to $g_0$. The dissipative current,
parallel to $\bE$, vanishes at the operating points  $E_x= \pm E_c$ .
There will be a non-zero Hall current in the $y$-direction. 
(c) Lyapunov function in the presence of a uniform disorder field $\bE_d$ in 
the x-direction.  
\label{gplot}}
\end{figure}

We now consider the clean system ($\bE_d=0$) in the regime of strong
microwave radiation at positive detuning, i.e., with microwave frequencies
slightly larger than the cyclotron harmonics $\omega=m\omega_c$, where $m=1,2,\ldots$.
Both DP and DF mechanisms produce a regime of negative longitudinal
conductivity around $E=0$, which implies a minimum of $g(E)$ at a
finite field of magnitude $E=E_c$. If the radiation is strong enough,
$E_c$ can  become the {\em global} minimum,and the Lyapunov density
describes a dynamical phase of spontaneous photogenerated fields. Then
$g$ can be expanded as
\be
g_0(\bE) = g_0(E_c) + \frac{1}{2}(E-E_c)^2 \sigma_c + \lambda |\nabla\cdot
\bE|^2\label{lyap-clean}, 
\ee
where the subscript 0 denotes the absence of disorder. A
representative $g$ and the corresponding dissipative current $j^d$ are
shown in Figs.~\ref{gplot}a and ~\ref{gplot}b. In order to
satisfy equipotential boundary conditions and the constraint $\nabla
\times \bE=0$, domain walls between different directions of $\bE$ must
form. The field derivative coefficient $\lambda \approx \sigma_c
l_{dw}^2$ implements the ultra-violet cutoff, giving a finite
domain-wall thickness scale $l_{dw}$. Domain walls give a positive
contribution to $G$ of order $\sigma_c E_c^2  l_{dw}$ per unit
length. In the absence of disorder, the system will simply minimize
this residual domain wall contribution, subject to aforementioned
constraints.

Local stability requires that  $g$ is concave near $E_c$. By
rotational symmetry $g_0$ is a `Mexican hat' function with a flat
valley along $|\bE|=E_c$. Away from domain walls, the fields are
`marginally' stable. The clean system has zero conductance in the
limit when the system size is much larger than $l_{dw}$.  The system
can accomodate a potential difference by moving domain walls and by
changing the direction of $\bE$ while keeping $|\bE|=E_c$, and keeping
$g_0$ constant everywhere away from domain walls.  The current density is
maintained at zero everywhere, except for possible contributions  that
vanish in the limit $l_{dw} \to 0$, which is a property of the clean
ZCS \cite{ZRS-macro}.

\section{Domain Structure in the Absence of Disorder}

\begin{figure}[!h]
\vspace{1mm}
\centerline{\epsfysize=3in\epsfbox{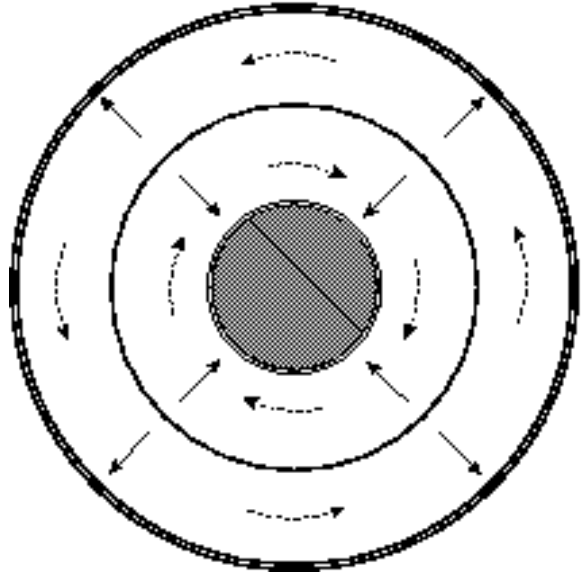}}
\vspace{1mm}
\caption{Domain structure for a circularly-symmetric Corbino geometry,
in the macroscopic limit, without disorder.  Solid circle, midway
between the inner and outer edges is the location of the domain wall
when there is no voltage difference between the two edges. Solid
arrows show the direction of the electrostatic field $\bE$, while
dotted arrows show the direction of the circulating Hall currents. An
alternate solution of the equations has the same position of the
domain wall, but reversed directions for the electric fields and Hall
currents. In a Corbino conductance measurement, one applies a voltage
difference between the inner and outer edges, and measures the
electrical current from one edge to the other.\label{corbino}}
\end{figure}

\begin{figure}[!h]
\vspace{1mm}
\centerline{\epsfysize=4.2in\epsfbox{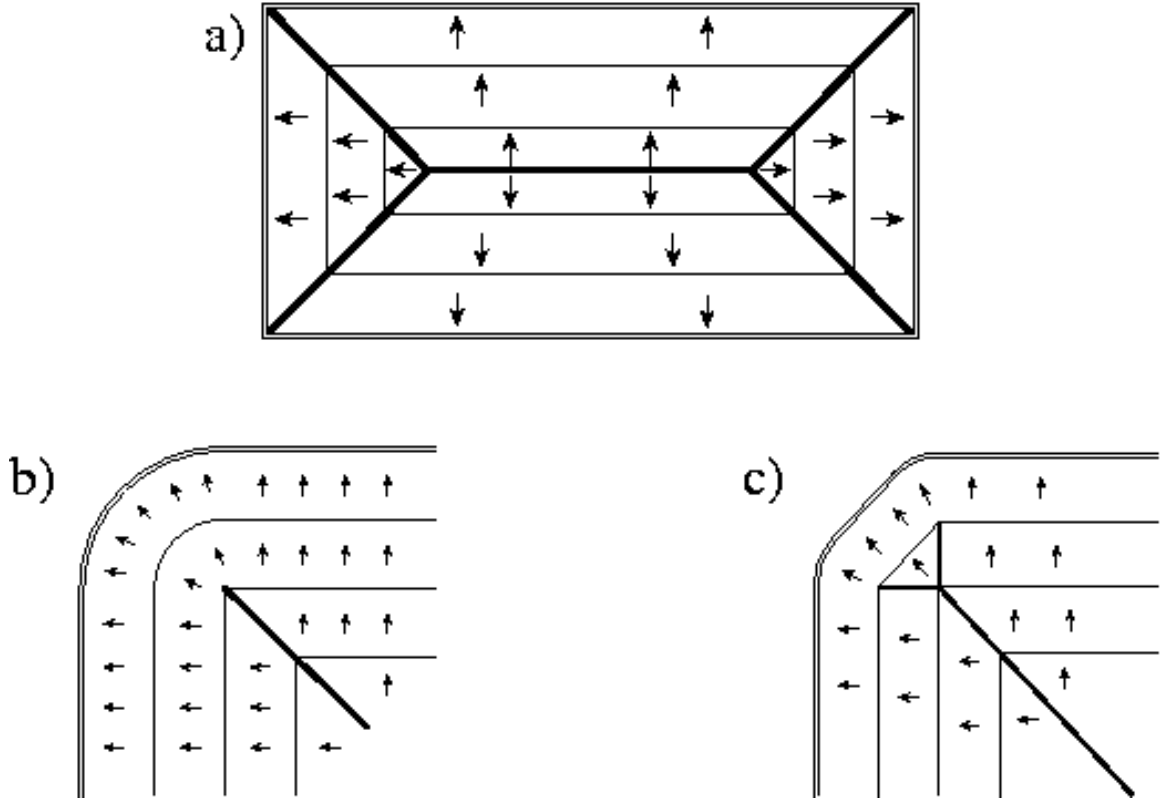}}
\vspace{1mm}
\caption{Other possible domain structures in a sample without
disorder.  Heavy solid lines show domain walls; light solid lines show
representative equipotential contours; and double lines show sample
boundaries.  Arrows show direction of electric field $\bE$.  Panel (a) shows structure for a rectangular sample. Panel (b)
shows how a domain wall can end  inside the sample, in the vicinity of
a curved corner.  Panel (c) shows how a domain wall can split
into two. 
\label{boundary}}
\end{figure}

Although the Lyapunov cost per unit length of a domain wall
depends on the precise angular difference of the orientations of the
electric fields on the two sides of it, the pattern of domain walls in
the simplest cases, in the absence of disorder,  can be understood as
an attempt to minimize the total domain wall length, subject to the
boundary constraints.  If the potential is a constant, which we may
take equal to zero,  everywhere along the boundary, we are led to the
trial solution:  
\be
\phi(\br) = E_c d(\br)
\label{phinodis}
\ee
where $d(\br)$ is the distance to the nearest boundary. The negative
of this solution, with $E_c$ replaced by $-E_c$, is an equally good
solution, with the same Lyaponov cost.  Domain walls for these
solutions occur along curves which are the loci of points that are
equidistant from two different points on the sample
boundary. Domain walls may end, or one domain wall may split into two,
at special points, which are equidistant from three different points
on the boundary. Some simple examples, illustrating these
possibilities, are given in Figs.~\ref{corbino} and~\ref{boundary}.

Clearly, the trial solution (\ref{phinodis}) satisfies the requirement
that $E=E_c$ everywhere except on the domain walls.  Moreover, the
requirement $E=E_c$ requires that in the absence of singularities, two
nearby equipotential contours must be separated from each other by a
constant distance $\delta$, equal to the difference in the potentials
divided by $E_c$.  Thus, if one starts from a point on the sample
boundary, the potential $\phi(\br)$ must be given by (\ref{phinodis})
or its negative at least until one encounters a domain wall.  If one
adds additional domain walls, one may construct solutions different
from these, but only by increasing the total length of the domain
walls and presumably increasing the Lyapunov cost.  Note that one
cannot simply displace a domain wall to one side or another without
violating the condition $E=E_c$, or introducing a discontinuity in
$\phi$ (i.e., an infinite electric field), which we do not allow.

In the circularly symmetric Corbino geometry, illustrated in Fig.~\ref{corbino},
we see that if there is no voltage difference between the inner and
outer edges, the domain wall must sit at radius $R=R_m$ where $R_m$ is
the arithmetic mean of the radii of the inner and outer edges. The
domain wall has a Lyapunov cost $\delta G = 2 \pi R g_w $, where $g_w$
is the cost per unit length of a 180-degree domain wall.  If the the
potential $\phi$ at the outer edge is increased by a small amount $V$,
while the potential at the inner edge is held constant, the position
$R$ of the domain wall must change by an amount $\delta R = \pm V / 2
E_c$, where the sign depends on which of the two solutions,
(\ref{phinodis}) or its negative, one is starting from.
Consequently, since $I=dG/dV$,  there should be a small but non-zero
photocurrent, $I = \pm  \pi g_w / E_c$ between the outer and inner
edges, at $V=0$.  

Non-zero $V$ breaks the perfect  degeneracy between the two steady
state solutions, as the solution with negative $\delta R$ has the
smaller value of  $G$, and  it has current  in the {\em opposite}
direction to the voltage drop. If the system can switch from one state
to another, the current will change direction discontinuously when $V$
passes through zero. In the absence of noise, or if V is changed
sufficiently rapidly, however, the system may be trapped in one state or the
other, in which case there will be no discontinuity at $V=0$. 

\section{Long Wavelength Disorder.}

 In an inhomogeneous system, there will generally be a non-zero
 electrostatic field, $\bE_d(\br) \equiv  - \nabla \phi_d(\br)$
 present in the thermal equilibrium state, with no microwave radiation
 or bias voltage. (Recall that it is the electrochemical potential,
 which is the sum of the electrostatic potential and the internal
 chemical potential, that is constant in equilibrium). We may ask how
 this disorder field will alter the ZCS. 

At weak disorder field, $|\bE_d|<<E_c$,  we expand the Lyapunov
density in  a region where $E\approx E_c$,
\be
g(\bE,\bE_d)= g_0(\bE) -   \sigma_1(E)  ~\bE \cdot \bE_d (\br)+\cO(E_d^2),
\label{Lyap}
\ee
which yields a current density
\be
\bj^d(\br) =  -\sigma_1 \bE_d + \frac{g_0'-\sigma_1'\bE\cdot\bE_d}{E}\bE,
\label{Curr}
\ee
where $X'\equiv \partial X/\partial E$, and the coefficient $\sigma_1$
depends on the microscopic mechanism.  A typical $g$ in the presense
of disorder is displayed in Fig.~\ref{gplot}c.  For a simple model based on the
displacement mechanism, we estimate that $\sigma_c$ and
$\sigma_1(E_c)$ should be close to the dark conductivity, while
$\sigma'_1 \ll\sigma_c/E_c$. Other models may have quite different
values of $\sigma_1$. 

Our use  of (\ref{Lyap}) assumes that the length scale of the disorder
field is greater than the domain wall thickness $l_{dw}$, which is the
same as our short wavelength cutoff.\cite{ZRS-disorder} In practice,
however, we will only be interested in fluctuations on length scales
much longer than this. In two dimensions, if the disorder potential
$\phi_d$ does not have correlations on a length scale much longer than
$l_{dw}$, and if  $E_d  \sigma_1 \ll E_c \sigma_c$, then the gain in
$G$ obtained by aligning $\bE$ with $\bE_d$ will be too small to
overcome the cost of introducing new domain walls.\cite{prl}  Then
even in the limit of a very large sample, the number and positions of
domain walls will be determined by the sample shape and boundary
conditions, just as if there were no disorder present.  By contrast,
if the disorder has correlations which fall off slowly at large
distances, then for the model given by (\ref{Lyap}), with a fixed
root-mean-square value of $E_d$, it will pay to introduce a network of
new domain walls, if the sample is sufficiently large, and we would
expect that local characteristics of the domain wall pattern will
become independent of the sample size. 

\subsection{Stability Conditions.}

Stability requires that the local conductivity tensor $\sigma^d$,
defined by (\ref{jdiss}), has non-negative eigenvalues at every
point. The lower (transverse)  eigenvalue is given by
\be
\sigma_{-}=\frac{g_0'-\sigma_1'\bE\cdot\bE_d}{E} + \cO(E_d)^2 \ge 0 \ ,
\label{cond-eig}
\ee
so  marginal stability occurs at $E=E_c+ \sigma_1'\bE_d\cdot\bE/\sigma_c$.

In a steady state, the normal current density across a domain wall
must be continuous. If $\hat\bn$ is the normal, and $\bE_1$ and
$\bE_2$ are the fields on the two sides, we find from (\ref{cond-eig})
and(\ref{Lyap}) that 
\be
\sigma_{-}(E_1) ~\bE_1\cdot\hat{\bn}=
\sigma_{-}(E_2) ~\bE_2\cdot\hat{\bn} ~+\cO(E_d^2)
\label{MS}
\ee
When the normal component of $\bE$ has opposite signs  on the two
sides, (as occurs in the clean limit), (\ref{MS}) can only be
satisfied for $\sigma_{-}(E_1)=\sigma_{-}(E_2)=\cO(E_d)^2$.  {\em This
  restricts the fields at the domain wall to be at their respective
  marginally stable values}.  Among the other consequences of these
equations, it can be shown that if a path along a set of domain walls 
forms a closed loop, the integral of the ``two-dimensional
disorder-charge density'' $q_d(\br) \equiv \nabla \cdot \bE_d /2\pi$,
over the area enclosed by the loop, must be equal to zero, up to
possible corrections of order $E_d^2$.\cite{prl} 

To a lowest order in $\frac{E_d}{E_c}$, the field $\bE$ will have
magnitude $E_c$ within each domain.  Coupled with the fact that the
parallel component of electric field is continuous across the domain wall,
this implies that the domain wall between any two domains is oriented
in such a way as to bisect the angle between the field lines of the
two domains.

Finally, we note that generically, the differential conductance of a
sample in the Corbino geometry can be obtained by solving for the
conductance of a linear system with local conductivity given by
$\sigma^d(\bE(\br))$, in series with resistive elements along the
domain walls, which arise from movement of the domain walls in
response to a variation in the applied bias $V$. There could also be
discontinuities in the current at discrete values of $V$, if the
system jumps discontinuously from one minimum of $G$ to another. We
find that for weak long-wavelength disorder,  the scale of the
macroscopic conductance is set by the domain-wall contribution. 

\subsection{One-Dimensional Disorder.}

The simplest example to consider is the case of one-dimensional
disorder, where $\phi^d$ is a function of $y$, independent of the
$x$-coordinate.  At wavelengths much larger than $l_{dw}$, if we do
not specify the total voltage drop $V$, the Lyapunov functional $G$ is
minimized  if the system breaks up into parallel domains, so that
$\bE$ is everywhere aligned with $\bE_d$, and $\bj^d = 0$. These
conditions determine $\bE(y)$  via (\ref{Curr}). They  are  consistent
with the boundary conditions for a rectangular Corbino geometry, 
which has periodic boundary conditions at $x=0$ and $x=L$, and
specified potentials at $y=0$ and $y=W$, provided that the voltage
difference $V$ satisfies 
\be V [I]=  \int_0^W dy (E^y(y)-E_d^y(y))  \  .  \label{1Dvoltage}
\ee
Thus, at strong  radiation intensity and zero current, we see that
that  the domain walls form precisely at maxima and minima of
$\phi_d$, at positions $y_i, i= 1,...N$,  where $E_d^y(y)$ changes
sign. For a general choice of $\phi_d(y)$, when no current is drawn
from the sample, there will be a non-zero photovoltage, whose value is
independent of the magnitude of $\phi_d$,  to lowest order, and is
determined by the positions $y_i$ :
\be V(j=0) =  E_c \left( (-1)^N W - 2 \sum_{i=1}^N (-1)^i  y_i \right) +\cO(E_d),
\label{PV}
\ee
where we have assumed that the odd values of $i$ correspond to maxima in $\phi_d$.

If there is a non-zero current $j^d_y$ (in the $y$-direction), the domain
walls will be displaced, leading to a change in voltage proportional
to $j^d_y$. For an appropriate choice of $j^d_y$, such that  the voltage
resulting from $j^d_y$ cancels the zero-current photovoltage, we can get
$V=0$.  Near $j^d_y = 0$, the differential conductivity $\sigma_{yy}$ is
given, to lowest order in $\phi_d$, by 
\be 
\frac{1}{\sigma_{yy}}  =\frac{2 E_c}{\sigma_1 L}  \sum_i \frac{1}{|\phi_d''(y_i)|} 
\equiv\frac{2E_c}{\sigma_1 \tilde{E}_d}.  \label{resis-1d}
\ee
If $P(E_d)$ is the probability distribution for $E_d$ at a random
point, it can be shown that $\tilde{E}_d^{-1} = P(0)$.
If $E_d$ is taken from a Gaussian distribution, then $\tilde{E}_d=(2
\pi)^{1/2}  E_d^{\rm{rms}}$, where $E_d^{\rm{rms}}$ is the root-mean
square value of $E_d$. If $\phi_d$ is a single sine wave, then
$\tilde{E} = 2^{1/2} \pi E_d^{\rm{rms}}$. The transverse differential
conductivity $\sigma_{xx}$, which can also be calculated using
(\ref{cond-eig}) and (\ref{Curr}), with $j^d_y=0$, is given, to a first order in $\phi_d$, by
$\sigma_{xx} = \sigma_d  <|E_d|> / E_c $, where $ <|E_d|>$ is the mean
value of $|E_d|$.  For a gaussian distribution, one has $\sigma_{xx}=
( 2 / \pi) \sigma_{yy}$, while for a single sine wave, $\sigma_{xx}= (
4 / \pi^2) \sigma_{yy}$. 

In Fig.~\ref{potential}a, we show potentials $\phi(y)$  for the cases   $j^d_y=0$
and  $V=0$, corresponding to a particular choice of the
one-dimensional disorder potential $\phi_d (y)$. 

\subsection{Two-Dimensional Disorder Potentials}
\subsubsection{Egg-carton potential}

Perhaps the simplest 2D choice for $\phi_d$  is the periodic separable
egg-carton potential, given by \be \phi_d = A \sin (k_x x) + B
\sin (k_y y) \  . \ee 
We consider here a finite system, a rectangle $L_x$ by $L_y$, with
periodic boudnary conditions on the current and the electric
fields. This, of course, is equivalent to considering an infinite system. For bias voltage $V=0$, we construct a trial solution for $\phi$ by
placing domain walls on the lines $x=\pi (2m+1)/2k_x$ and $y=\pi (2n+1)/2k_y$, for integer
$n$ and $m$.  Our trial solution will have a constant electric field in each
rectangular domain: 
\be \bE = E_c(\pm\hat{x}  \cos \theta_0  \pm\hat{y} \sin \theta_0 ) \
.  \label{ensatz} \ee
Here, $\theta_0$ is the angle between the field and the y axis.
Combining equations (\ref{Lyap}) and  (\ref{ensatz}), we see that, to
a linear order, minimization of $G$ is equivalent to maximization of $\int d^2x  \bE \cdot \bE_d . \label{linear}$ The value of $G$ is minimized with the choice $ \theta_0 = \arctan(Ak_x/Bk_y) $. The domain walls satisfy the neutrality condition that
$\int q_d(\br) d^2r=0$ in each domain, and the electric field is the
gradient of a continuous potential $\phi$ as required. 

In the
symmetric case, $k_x=k_y, A=B \equiv E_d^0/ (2^{1/2}k_x), $ the domain
walls form a square array, and $\theta_0 = \pi/4$. We have carried out
a numerical minimization of $G$, and have plotted the ``two dimensional
charge density'', $q(\br) \equiv  \nabla \cdot \bE/2\pi$, for  the symmetric egg-carton
potential, for zero bias voltage in Fig.~\ref{walls}a. Domain walls,
appearing here as  line singularities, form a square array. Evidently,
the numerical solution we obtained is very close to the variatonal one above.

We now construct a variational solution for the case where there is a
nonzero voltage drop $V$ along $x$-direction. Assuming that the electric field is still
constant within each domain, we allow the centers of domain walls that
are perpendicular to the voltage drop to
shift by amount $\pm \delta x$ and rigidly rotate walls through an angle
$\pm \delta \theta$. The electric fields in the various domains will
now lie in the directions $\pm\left(\theta_0+\delta\theta\right)$ and
$\pm\left(\pi+\theta_0-\delta\theta\right)$, so that the domain walls still bisect
the angle formed by the field lines in the neighboring
domains. In Fig.~\ref{walls}b, we plot the (numerically obtained) ``two dimensional charge density'',
for a bias voltage corresponding to an average electric field
equal to  $0.25 E_c$ in the $x$ direction. The numerical solution
is very much like the variational one just described in that
domain walls perpendicular to voltage drop are shifted and tilted into
a herringbone pattern, while the walls along the voltage drop are
essentially unchanged. 

The variational solution above can be used to evaluate conductivity
$\sigma_{xx}$.  For a voltage drop $V$ along $x$-direction, we
have
\be \frac{2\pi V}{E_c L_x
  k_x}=4 \delta x \sin \theta_0+\frac{2\pi}{k_x}\delta \theta \cos \theta_0. \label{vdrop}\ee
Lyapunov functional $G$ will then be
\be G(V)=G_0+G_1\left(\frac{1}{2} \left(k_x \delta x\right)^2  \sin^2
\theta_0+\delta\theta^2\left(\frac{1}{2}+\frac{\pi^2}{24}\left(\frac{k_x}{k_y}\right)^2
\sin^2\theta_0\right)\right), \ee where $G_0$ is the Lyapunov
``energy'' for the case of $V=0$ and $G_1= \frac{2\sigma_1 E_c L_x
  L_y}{\pi}\sqrt{\left(A k_x\right)^2+\left(B k_y\right)^2}$. Using
Eq.\ref{vdrop}, we mininimize the quadratic form above. By taking derivatives of the resulting expression with
respect $V$, as prescribed by Eq.\ref{conduct}, we obtain the
following expression for $\sigma_{xx}=\frac{L_x}{L_y}G''(0)$:
\be
\sigma_{xx}=\frac{\pi \sigma_1}{2 }~\frac{\sqrt{(A k_x)^2+(B
    k_y)^2}}{E_c}~\frac{1+\frac{\pi^2}{12}\sin^2 \theta_0
  \left(\frac{k_x}{k_y}\right)^2}{1+\frac{\pi^2}{12}\sin^2 \theta_0
  \left(\frac{k_x}{k_y}\right)^2+\frac{\pi^2}{4} \cos^2\theta_0}. \ee For the symmetric case, this reduces to $\sigma_{xx} = 0.83 \sigma_1
 E_d^0/ E_c$. As in the 1D case, the conductance is of order $\sigma_1
E_d^{\rm{rms}}   / E_c$, reflecting the pinning effect of the
``disorder  potential'' $\phi_d$. 

Numerical solutions, referred to above, have been obtained by
numerically minimizing the Lyapunov functional. We employ a
MATLAB-based routine to minimize the functional, which we
discretize on a triangular lattice (typically containing around 14,000
points) in order reduce any perturbations arising from lattice
anisotropy.  The potential $\phi(\br)$ is subject periodic boundary
conditions in the $y$-direction, and is required to change by a chosen
bias voltage $V$ in the $x$-direction.  Having started with a random
initial guess for $\phi(\br)$, the code allows the potential to relax to a
local minimum.  The robustness of a minimum, i.e. its ``globality'',
can be ascertained by starting with a different initial guess and/or
giving the solution we arrive at random kicks and plugging that in for
the next initial guess.  As already mentioned, for the case of
egg-carton potential, we find that the variational solutions above are
quite close to the ones we arrive at via numerics.  

\subsubsection{Disordered case}
\begin{figure}[!h]
\vspace{1mm}
\centerline{\epsfysize=3in\epsfbox{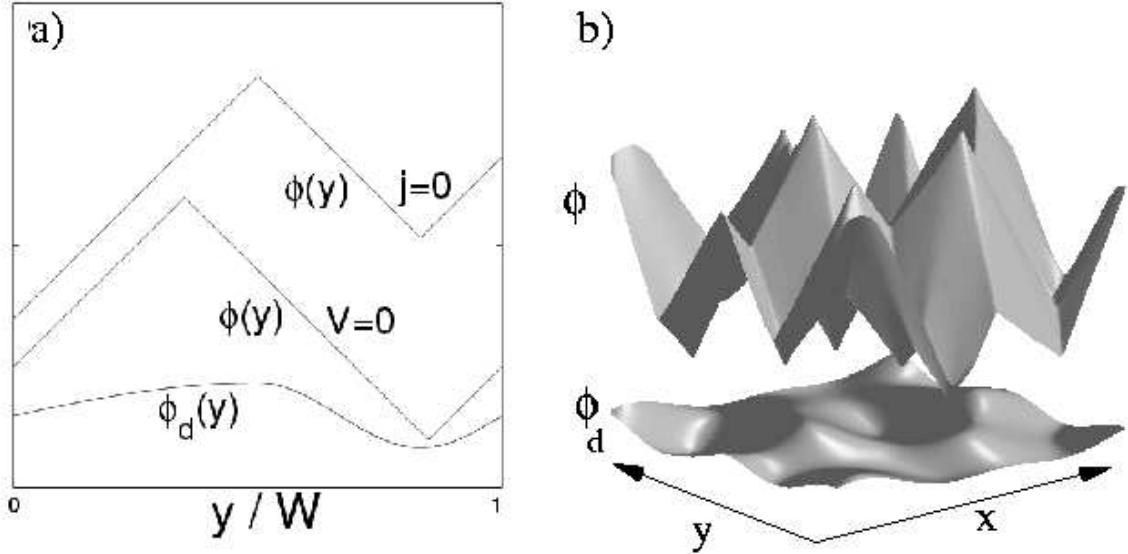}}
\vspace{1mm}
\caption{Examples of a disorder potential $\phi_d (\br)$ and the
resulting self-consistent potential $\phi(\br)$. Different vertical
scales were used for $\phi$ and $\phi_d$.  Panel (a) shows an
example of one-dimensional disorder, where $\phi_d$ and $\phi$ depend
only on the coordinate $y$.  Solutions are shown for two different
boundary conditions, corresponding to zero net voltage drop and to
zero longitudinal current flow. Panel (b) shows  $\phi(\br)$
resulting from a particular  two-dimensional disorder potential
$\phi_d$,  indicated in the lower plot. The disorder potential
contains 20 Fourier components, whose amplitudes were chosen from
Gaussian distributions with variance independent of the wave-vector
$\bf {k}$.  Periodic boundary conditions were imposed on $\phi$ and
$\phi_d$.
\label{potential}}
\end{figure}

\begin{figure}[!h]
\vspace{1mm}
\centerline{\epsfysize=2in\epsfbox{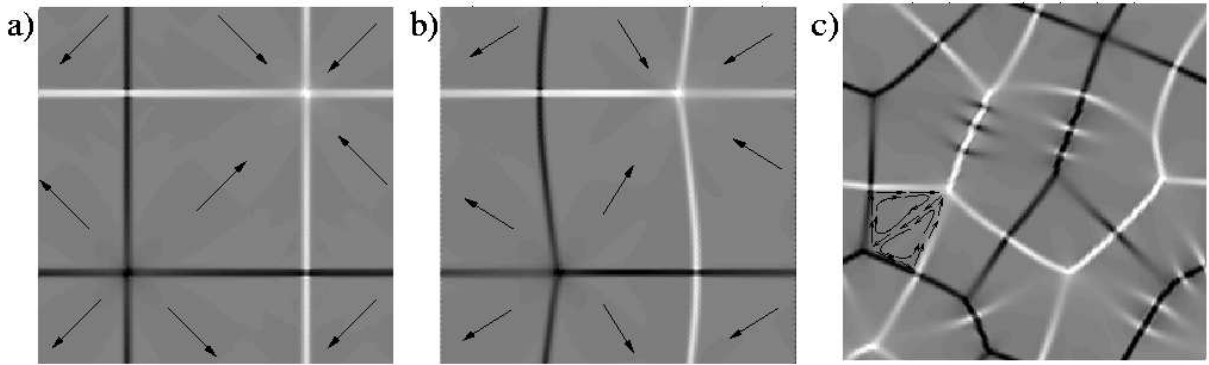}}
\vspace{1mm}
\caption{Gray-scale plots of the ``two-dimensional charge density'' 
$q(\br) \equiv  \nabla \cdot \bE /2\pi$, for three examples discussed
in the text.  Domain walls are visible as line singularities in the
charge density.  Panel (a) shows $q(\br)$, in  one unit cell,
when $\phi_d$ is the periodic symmetric egg-carton potential,  $\phi_d
\propto \cos x + \cos y$, with zero net voltage drop across the
sample.  The arrows show the direction of electric field within each
domain. Panel (b) shows $q(\br)$ for the same choice of
$\phi_d$, with a non-zero voltage bias, corresponding to an average
electric field $E = 0.25 E_c$ in the $x$-direction.  Once again, the
arrows show the direction of field within each domain. Panel (c)
shows $q(\br)$ for the more complicated example of $\phi$ and $\phi_d$
illustrated in Fig.~\ref{potential}b. Arrows here show circulating
dissipative currents in one domain. 
\label{walls}}
\end{figure}

The numerical routine allows us to obtain solutions for disorder
potentials much more complicated than the simple egg-carton
potential. For example, in Fig.~\ref{potential}b,  we display the self-consistent potential $\phi(\br)$ for
the case of $\phi_ d$ containing 20 Fourier
components, chosen from a gaussian distribution with
$<|\phi_d({\bf{k}}|^2> $ independent of ${\bf{k}}$.  The disorder
potential $\phi_d$ is also shown. In Fig.~\ref{walls}c, we plot the
``two-dimensional charge density'', proportional to $\nabla \cdot
\bE$, corresponding to this solution.  Domain walls again appear as
line singularities in the charge.  Although one might expect
frustration to reduce the conductance in complicated potentials, the
conductivity in this case was found to be similar to that for an
egg-carton potential with the same value of $E_d^{\rm{rms}} $. 

In both the egg carton potential and the more complicated potential of
Fig.~\ref{potential}b, there is frustration in the minimization of $G$. The field
$\bE$ is unable to align perfectly  with $\bE_d(\br)$, because of the
conflicting requirement of $E \ge E_c$, which arises from stability,
and $\nabla \times \bE=0$. Because of this, the dissipative current
$j^d$ is generally  non-zero, of order $\sigma_1 E_d$, leading to
circulating dissipative currents within each domain.  These are
indicated  by arrows for one domain in Fig.~\ref{walls}c.   The much larger
circulating Hall currents, of magnitude $\sigma^H E_c$, are not shown
in the figures.  

\section {Discussion.} 

In  this paper, we have explored the effects of sample boundaries and
long-wavelength potential disorder, on the domain pattern and
conductance of the microwave-induced zero-resistance state, within a
simple model.  In particular, our model, with a constant Hall
conductance, has the simplifying feature that   it has a Lyapunov
functional. In considering the effects of disorder, we assumed a
linear coupling to the disorder-induced electrostatic field $\bE_d$.
Furthermore, while we consider non-linear effects due to the
self-consistent electric field $\bE$, we do not consider
non-linearities due to changes in the local electron-density itself,
which may result from the electric fields  of the domains. (We might
expect density fluctuations to be small, while potential fluctuations
are large, if the system is far from any screening conductors.)  For
more complicated systems, where there is no Lyapunov functional, we do
not even know whether the system will reach a time-independent steady
state, in the presence of strong microwave radiation.    

Even for our simple model, with its Lyapunov functional $G$, there are
many questions to be answered. What actually is the behavior of a
system with two-dimensional disorder, in the limit of a large system
size? Will there be a large number of metastable steady states, as in
a glass, with associated hysteresis in the observed current/voltage
curve?  If we assume that for  all values of the potential difference
$V$ between the inner and outer edge, in  the Corbino geometry, the
system can reach the state with lowest value of  $G$, will the
conductance be finite in the limit of large system size, with
magnitude of disorder held fixed, or will it, perhaps, go to zero in
this limit?

Experimental work in the field is also in its infancy.  The
self-consistent electrostatic potential $\phi(\br)$ generated by the
domains can in principle be measured with minimal disturbance of the
sample, using, for example, a single electron transistor device.  Such
a method has been used successfully to explore various aspects of the
microscopic properties of the integer and fractional quantum Hall
effect\cite{ilani}, and we may expect important information about domain
structures to  come from such measurements in the future. 

We remark that Alicea et al.\cite{alicea} have shown that a Lyapunov
functional exists, rather generally, very close to the dynamical phase
transition, where the microwave power is such that the longitudinal
conductivity first becomes negative at $E=0$.  However, the form of
their functional is different from the one employed in the present
paper.  We are interested here in systems under strong microwave
irradiation, far from the phase transition, so the Lyapunov functional
of Alicea et al does not apply. 

{\bf Acknowledgments}. We thank M. Lukin, S. Fishman,  E. Meron, and
 Y. Yacoby for helpful discussions.  Work was supported in part by the
 Harvard Center for Imaging and Mesoscale Structures,  NSF grant
 DMR02-33773, US-Israel Binational Science Foundation, Minerva
 Foundation and the DFG priority program on Quantum Hall Systems
 YA111/1-1.  One of the authors, AA, would also like to acknowledge
 the hospitality of Aspen Center for Physics.  


\clearpage
\clearpage

\clearpage
\clearpage

\clearpage
\clearpage

\clearpage
\clearpage

\clearpage
\clearpage

\end{document}